\documentclass[final,3p,times,12pt]{elsarticle}
\usepackage{graphicx,color,amsmath,amssymb}
\usepackage{bm}
\usepackage[colorlinks=true, pdfstartview=FitV, linkcolor=red, citecolor=blue, urlcolor=blue]{hyperref}

\advance\voffset -0.2in
\biboptions{sort&compress}

\begin{document}

\begin{frontmatter}
\title{Multi-particle eccentricities in collisions dominated by fluctuations}

\author[agh,bnl]{Adam Bzdak}
\ead{bzdak@fis.agh.edu.pl}

\author[bnl,wmu]{Vladimir Skokov}
\ead{vskokov@quark.phy.bnl.gov}

\address[agh]{AGH University of Science and Technology, Faculty of Physics and Applied Computer Science, \\30-059 Krak\'ow, Poland}
\address[bnl]{RIKEN BNL Research Center, Brookhaven National Laboratory, 
Upton, NY 11973, USA}
\address[wmu]{Department of Physics, Western Michigan University, Kalamazoo, MI 49008, USA}

\begin{abstract}
We compute analytically the multi-particle
eccentricities, $\epsilon_{m}\{2n\}$, for systems dominated by fluctuations, 
such as proton-nucleus collisions at the Large Hadron Collider. 
In particular, we derive a general
relation for $\langle \epsilon_{2}^{2n} \rangle$. We further discuss the relations between various
multi-particle eccentricities and demonstrate that $\epsilon
_{2}\{2\}>\epsilon _{2}\{4\}\simeq \epsilon_{2}\{6\}\simeq
\epsilon_{2}\{8\} $, in agreement with recent numerical
calculations in a Glauber model.
\end{abstract}
\end{frontmatter}


\section{Introduction}

Recent measurements of high multiplicity proton-proton (p+p) and
proton-nucleus (p+A) collisions at the Large Hadron Collider (LHC) revealed
an unexpected enhancement of the two-particle correlation function at
small azimuthal angles and large separation in rapidity \cite%
{Khachatryan:2010gv,CMS:2012qk,Abelev:2012ola,Aad:2012gla}. This effect was
also seen at the Relativistic Heavy Ion Collider (RHIC) in deuteron-gold (d+Au) and helium-gold ($^3$He+Au) collisions \cite{Adare:2013piz,Adare:2015ita}.

The same correlation pattern was previously observed in nucleus-nucleus
(A+A) collisions at RHIC~\cite{RidgeatRHIC}. In nucleus-nucleus collisions
the azimuthal angle correlation function and its Fourier harmonics are well
described by the relativistic \textit{hydrodynamics}, an effective theory of
long wave excitations in a strongly coupled system \cite{Florkowski:book}.
Using viscous hydrodynamics, the ratio of the shear viscosity over the entropy is
found to be surprisingly small (see, e.g., Ref.~\cite{Song:2008hj}) and close to
the conjectured lowest bound for a strongly interacting system~\cite%
{Son:2007vk}.

Nucleus-nucleus collisions are immensely complicated, due to multi-particle
rescattering, the possible formation of a thermal system, and subsequent
collective evolution. They do not offer a direct possibility to study
initial state effects. It was expected that elementary p+p and p+A
collisions are dominated by the initial state effects and thus their
behavior can be studied and described by quantum chromodynamics (QCD) at
weak coupling. However, due to the high densities of partons, the effects of gluon
saturation must be taken into account. This is done in the
framework of the \textit{Color Glass Condensate} (CGC)~\cite{Gelis:2010nm},
an effective description of a hadron at asymptotically high energy in the
regime of weakly coupled QCD. Although at present there is no compelling
experimental evidence indicating that the CGC is an appropriate tool to
interpret hadronic collisions at the LHC energies, there are 
attempts to describe the azimuthal angle correlation functions in p+p and
p+A collisions at the LHC~\cite{Dusling:2013oia}, see also Refs.~\cite{Kovchegov:2012nd,Kovner:2012jm,Schenke:2015aqa}, and very recent development in Ref. \cite{DS}, which in particular showed that conventional CGC used in Ref.~\cite{Dusling:2013oia} is incompatible with the experimental data at high multiplicity.
Recently, hydrodynamics
was applied to p+A collisions \cite{Bozek:2011if, Bzdak:2013zma,
Bozek:2013yfa, Qin:2013bha, Shuryak:2013ke,Kozlov:2014fqa} with a reasonably good fit to
the data. This success however does not solve several conceptual theoretical
problems, such as how quickly thermalization occurs, whether the initial conditions are boost invariant, and
many other effects, which are implicitly assumed when hydrodynamics is applied.
Very recently some of these problems were addressed in the AdS/CFT framework 
in Refs.~\cite{Kalaydzhyan:2014zqa, Iatrakis:2015rga}. Finally, a multi-phase transport model (AMPT) \cite{Lin:2004en} was recently compared with the experimental data in p+p \cite{Ma:2014pva}, p+A \cite{Bzdak:2014dia} and d+Au \cite{Koop:2015wea} interactions. Possible origin of the anisotropies  within  the AMPT
model are discussed in Ref. \cite{He:2015hfa}. 

In summary, at present we have two general approaches for the high multiplicity p+p and
p+A collisions at the LHC energy: models of 
strongly interacting medium (hydrodynamics, AdS/CFT, cascade) and a rival effective theory of QCD at high
energies in the weakly interacting regime, the CGC.  Several observables and ideas
were recently put forward to single out an appropriate language to describe
phenomena in these collisions \cite{Bzdak:2013zla,Bozek:2013sda,Coleman-Smith:2013rla,Basar:2013hea}.

The motivation for this short note is the observation of Ref.~\cite%
{Bzdak:2013rya}, where the authors showed that the initial eccentricities of
the interaction region in p+A and A+A interactions form a peculiar
hierarchy, namely, the eccentricities computed with the two and higher
number of particles satisfy the following relation: 
\begin{equation}
\epsilon _{2}\{2\}>\epsilon _{2}\{4\}\simeq \epsilon _{2}\{{6\}}\simeq
\epsilon _{2}\{8\}\simeq \ldots   \label{rel}
\end{equation}%
This relation was also verified in Ref. \cite{Yan:2013laa}, where its origin
was attributed to a power law distribution of $\epsilon_{2}$.

Equation (\ref{rel}) has serious phenomenological implications. First,
if the same hierarchy is observed for the Fourier coefficients of the
azimuthal correlation function in p+A collisions, it would indicate that the
azimuthal correlation of hadrons is determined by the geometry of the
initial state. 
This favors approaches where the initial geometry is translated into momentum
space from collective effects, such as in hydrodynamics.
Current treatments of the CGC are independent of the geometry,
so that equivalent hierarchy for the Fourier coefficients of the azimuthal correlation 
function is not apparent. 

In this paper, we extend the numerical results of Ref.~\cite{Bzdak:2013rya}
by analytically computing various eccentricities in a system dominated by
fluctuations, e.g., p+A. We provide compact analytical expressions that can
be used for further analysis.

\section{Derivation of analytical results for eccentricities}

Suppose we distribute $N$ points on a plane. Let the distribution be 
$P(r,\phi )$, where $r$ and $\phi $ are the polar coordinates of the points.\footnote{The distribution 
$P(r,\phi )$ is normalized to unity.}
The ellipticity squared in a given event is defined as \cite%
{Alver:2006wh,Teaney:2010vd} 
\begin{equation}
\epsilon _{2}^{2}=\frac{\left[ \sum_{i=1}^{N}r_{i}^{2}\cos (2\phi _{i})%
\right] ^{2}+\left[ \sum_{i=1}^{N}r_{i}^{2}\sin (2\phi _{i})\right] ^{2}}{%
\left( \sum_{i=1}^{N}r_{i}^{2}\right) ^{2}}=\frac{\left(
\sum_{k=1}^{N}r_{k}^{2}e^{i2\phi _{k}}\right) \left(
\sum_{l=1}^{N}r_{l}^{2}e^{-i2\phi _{l}}\right) }{\left(
\sum_{i=1}^{N}r_{i}^{2}\right) ^{2}}.  \label{e22}
\end{equation}

The goal of this brief note is to calculate $\left\langle \epsilon
_{2}^{2n}\right\rangle $ analytically for an arbitrary value of $n$, where $%
\left\langle ...\right\rangle $ denotes the average over many events. All
our results can be immediately generalized to, e.g., triangularity by
replacing $r^{2}\rightarrow r^{3}$ etc.\footnote{%
This depends on the definition of triangularity and higher eccentricities.
If we define them with $r^{2}$ than our results hold for all eccentricities.}
In this calculation we make two assumptions which, as we argue below, are
well justified. First, calculating Eq. \eqref{e22} we assume that $%
\left\langle A/B\right\rangle =\left\langle A\right\rangle /\left\langle
B\right\rangle $.\footnote{We found that this approximation reproduces the exact result with a Gaussian distribution $P(r)$, see Ref. \cite{Yan:2013laa}.} In this case
\begin{equation}
\left\langle \epsilon _{2}^{2n}\right\rangle \equiv \frac{\left\langle
\epsilon _{2}^{2n}\right\rangle _{\mathrm{num}}}{\left\langle \epsilon
_{2}^{2n}\right\rangle _{\mathrm{deno}}}=\frac{\left\langle \left(
\sum_{k=1}^{N}r_{k}^{2}e^{i2\phi _{k}}\right) ^{n}\left(
\sum_{l=1}^{N}r_{l}^{2}e^{-i2\phi _{l}}\right) ^{n}\right\rangle }{%
\left\langle \left( \sum_{i=1}^{N}r_{i}^{2}\right) ^{2n}\right\rangle }.
\label{e2n2}
\end{equation}%
To simplify our notation we introduce $\left\langle \epsilon
_{2}^{2n}\right\rangle _{\mathrm{num}}$ and $\left\langle \epsilon
_{2}^{2n}\right\rangle _{\mathrm{deno}}$ to denote the numerator and the
denominator of Eq.~\eqref{e2n2}.

In the following we neglect the recentering correction, i.e., the coordinate
system is not shifted to the center of mass. We expect this correction to
modify slightly our results only for small $N$. We will come back to this
point later in this Section, were we compare our analytical calculations with Monte
Carlo (MC) simulations.

To simplify equations we introduce the following notation 
\begin{eqnarray}
\mathcal{D}r &=&dr_{1}\ldots dr_{N}r_{1}\ldots r_{N},\quad  \\
\mathcal{D}\phi  &=&d\phi _{1}\ldots d\phi _{N},\quad  \\
\mathcal{P}(r,\phi ) &=&P(r_{1},\phi _{1})...P(r_{N},\phi _{N}),
\end{eqnarray}%
where in the last equation we explicitly assume that all the points are
sampled independently. The average over many events  is thus 
\begin{equation}
\langle A \rangle  = \int \mathcal{D} r \mathcal{D} \phi \mathcal{P}(r,\phi)  A \;.
\end{equation}

The denominator can be straightforwardly computed: 
\begin{eqnarray}
\left\langle \epsilon _{2}^{2n}\right\rangle _{\mathrm{deno}} &=&\int 
\mathcal{D}r\mathcal{D}\phi \mathcal{P}(r,\phi )\left(
\sum\nolimits_{i=1}^{N}r_{i}^{2}\right) ^{2n}  \notag \\
&=&\int \mathcal{D}r\mathcal{D}\phi \mathcal{P}(r,\phi )\lim_{x\rightarrow 0}%
\frac{d^{2n}}{dx^{2n}}e^{-x\sum_{i=1}^{N}r_{i}^{2}}  \notag \\
&=&\lim_{x\rightarrow 0}\frac{d^{2n}}{dx^{2n}}\left( \int d\phi rdrP(r,\phi
)e^{-xr^{2}}\right) ^{N}  \notag \\
&=&\lim_{x\rightarrow 0}\frac{d^{2n}}{dx^{2n}}\left( \int
rdrP_{r}(r)e^{-xr^{2}}\right) ^{N}  \notag \\
&=&\lim_{x\rightarrow 0}\frac{d^{2n}}{dx^{2n}}\left\langle
e^{-xr^{2}}\right\rangle ^{N}.  \label{e2n2deno}
\end{eqnarray}%
It is worth emphasizing that Eq.~\eqref{e2n2deno} is valid for an arbitrary $%
P(r,\phi )$.

Following the same procedure, the numerator of Eq.~\eqref{e2n2} can be written in the form   
\begin{eqnarray}
\left\langle \epsilon _{2}^{2n}\right\rangle _{\mathrm{num}} &=&\int 
\mathcal{D}r\mathcal{D}\phi \mathcal{P}(r,\phi )\left(
\sum\nolimits_{k=1}^{N}r_{k}^{2}e^{i2\phi _{k}}\right) ^{n}\left(
\sum\nolimits_{l=1}^{N}r_{l}^{2}e^{-i2\phi _{l}}\right) ^{n}  \notag \\
&=&\int \mathcal{D}r\mathcal{D}\phi \mathcal{P}(r,\phi )\lim_{x\rightarrow 0}%
\frac{d^{n}}{dx^{n}}\lim_{y\rightarrow 0}\frac{d^{n}}{dy^{n}}\exp \left(
-x\sum\nolimits_{k=1}^{N}r_{k}^{2}e^{i2\phi
_{k}}-y\sum\nolimits_{l=1}^{N}r_{l}^{2}e^{-i2\phi _{l}}\right)  \notag \\
&=&\lim_{x\rightarrow 0}\frac{d^{n}}{dx^{n}}\lim_{y\rightarrow 0}\frac{d^{n}%
}{dy^{n}}\left( \int d\phi rdrP(r,\phi )\exp \left( -xr^{2}e^{i2\phi
}-yr^{2}e^{-i2\phi }\right) \right) ^{N}  \notag \\
&=&\lim_{x\rightarrow 0}\frac{d^{n}}{dx^{n}}\lim_{y\rightarrow 0}\frac{d^{n}%
}{dy^{n}}\left\langle \exp \left( -xr^{2}e^{i2\phi }\right) \exp \left(
-yr^{2}e^{-i2\phi }\right) \right\rangle ^{N}.  \label{e2n2nom}
\end{eqnarray}

This equation can be used to express $\left\langle \epsilon
_{2}^{2n}\right\rangle _{\mathrm{num}}$ through $\left\langle r^{k}e^{im\phi
}\right\rangle $ for a general function $P(r,\phi )$.\footnote{%
For example, $\left\langle \epsilon _{2}^{2}\right\rangle _{\mathrm{num}%
}=N(N-1)\left\langle r^{2}e^{i2\phi }\right\rangle \left\langle
r^{2}e^{-i2\phi }\right\rangle +N\left\langle r^{4}\right\rangle $.} Similar
problem was extensively studied in the literature up to $n=2$ \cite%
{Bhalerao:2011bp,Alver:2008zza}.  Our Eq. (\ref{e2n2nom}) allows to derive an exact
relation for an arbitrary value of $n$. However, in the present paper we are
only interested in the  system dominated by fluctuations, that is, we neglect the 
$\phi $ dependence in $P(r,\phi )$ and assume that $P(r,\phi )=P_{r}(r)$.
Expanding exponents in Eq.~\eqref{e2n2nom} we obtain%
\begin{eqnarray}
\left\langle \epsilon _{2}^{2n}\right\rangle _{\mathrm{num}}
&=&\lim_{x\rightarrow 0}\frac{d^{n}}{dx^{n}}\lim_{y\rightarrow 0}\frac{d^{n}%
}{dy^{n}}\left\langle \sum\nolimits_{k=0}^{\infty
}\sum\nolimits_{l=0}^{\infty }\frac{(-xr^{2})^{k}(-yr^{2})^{l}}{k!l!}%
e^{i2\phi (k-l)}\right\rangle ^{N}  \notag \\
&=&\lim_{x\rightarrow 0}\frac{d^{n}}{dx^{n}}\lim_{y\rightarrow 0}\frac{d^{n}%
}{dy^{n}}\left\langle \sum\nolimits_{k=0}^{\infty }\frac{(xyr^{4})^{k}}{k!k!%
}\right\rangle ^{N}  \notag \\
&=&\lim_{x\rightarrow 0}\frac{d^{n}}{dx^{n}}\lim_{y\rightarrow 0}\frac{d^{n}%
}{dy^{n}}\left\langle I_{0}(2r^{2}\sqrt{xy})\right\rangle ^{N}  \notag \\
&=&\lim_{x\rightarrow 0}\frac{d^{n}}{dx^{n}}\lim_{z\rightarrow 0}x^{n}\frac{%
d^{n}}{dz^{n}}\left\langle I_{0}(2r^{2}\sqrt{z})\right\rangle ^{N}  \notag \\
&=&n!\lim_{z\rightarrow 0}\frac{d^{n}}{dz^{n}}\left\langle I_{0}(2r^{2}\sqrt{%
z})\right\rangle ^{N} ,  \label{e2n2noms}
\end{eqnarray}
where $I_{0}(x)$ is the modified Bessel function. Again, Eq. \eqref{e2n2noms} can be 
used to derive a relation between $%
\left\langle \epsilon _{2}^{2n}\right\rangle _{\mathrm{num}}$ and $%
\left\langle r^{k}\right\rangle $. We will present appropriate expressions
in Section \ref{sec:general-relations}.

The final result of our computations is given by 
\begin{equation}
\left\langle \epsilon _{2}^{2n}\right\rangle =\frac{n!\lim\limits_{z%
\rightarrow 0}\frac{d^{n}}{dz^{n}}\left\langle I_{0}\left( 2\sqrt{z}%
r^{2}\right) \right\rangle ^{N}}{\lim\limits_{z\rightarrow 0}\frac{d^{2n}}{%
dz^{2n}}\left\langle e^{-r^{2}z}\right\rangle ^{N}}.  \label{e2n2f}
\end{equation}

It would be interesting and pedagogical to demonstrate the effectiveness of
this result for a simple case of a Gaussian distribution.

\subsection{Gaussian distribution}

Equation~\eqref{e2n2f} can be computed analytically for a normalized
Gaussian distribution, $P_{r}(r)=2e^{-r^{2}/\sigma ^{2}}/{\sigma ^{2}}$. In
this case%
\begin{equation}
\left\langle e^{-r^{2}z}\right\rangle =2\int_{0}^{\infty }\frac{rdr}{\sigma
^{2}}e^{-r^{2}/\sigma ^{2}}e^{-r^{2}z}=\frac{1}{1+z\sigma ^{2}}
\end{equation}%
and%
\begin{equation}
\left\langle I_{0}(2r^{2}\sqrt{z})\right\rangle =2\int_{0}^{\infty }\frac{rdr%
}{\sigma ^{2}}e^{-r^{2}/\sigma ^{2}}I_{0}(2r^{2}\sqrt{z})=\frac{1}{\sqrt{%
1-4z\sigma ^{4}}}.
\end{equation}

\begin{figure}[h]
\begin{center}
\includegraphics[scale=0.43]{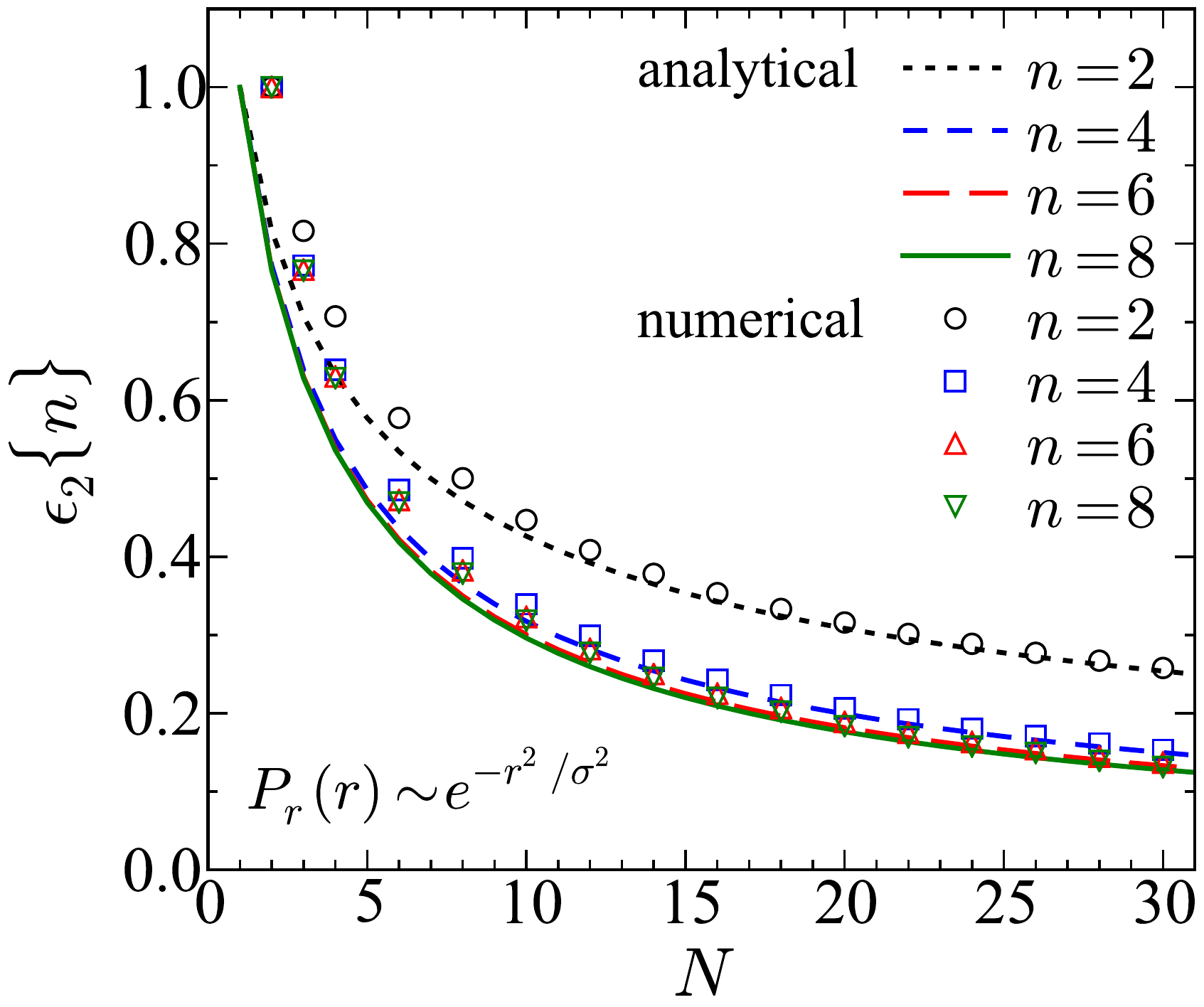}
\end{center}
\caption{The $n$-particle ellipticities $\protect\epsilon _{2}\{n\}$, $%
n=2,4,6,8$ for various numbers of independent points, $N$, calculated
analytically, Eqs. (\ref{e22g}-\ref{e28g}), compared with the Monte Carlo
(MC) calculations (open symbols). As we checked, the difference between MC
and Eqs. (\ref{e22g}-\ref{e28g}) comes solely from the recentering
correction which we neglect in our analytical calculations.}
\label{fig:1}
\end{figure}

Substituting into Eq.~\eqref{e2n2f} and differentiating with respect to $z$ we obtain
\begin{eqnarray}
\left\langle \epsilon _{2}^{2n}\right\rangle _{\mathrm{num}}
&=&n!4^{n}\sigma ^{4n}\frac{N}{2}\left( \frac{N}{2}+1\right) ...\left( \frac{%
N}{2}+n-1\right) , \\
\left\langle \epsilon _{2}^{2n}\right\rangle _{\mathrm{deno}} &=&\sigma
^{4n}N(N+1)...(N+2n-1)
\end{eqnarray}%
and finally%
\begin{equation}
\left\langle \epsilon _{2}^{2n}\right\rangle =\frac{n!4^{n}\frac{N}{2}\left( 
\frac{N}{2}+1\right) ...\left( \frac{N}{2}+n-1\right) }{N(N+1)...(N+2n-1)}. \label{e22ng}
\end{equation}%
As expected $\left\langle \epsilon _{2}^{2n}\right\rangle $ does not depend
on the width of the distribution, $\sigma $. Using Eq. (\ref{e22ng}) various cumulants 
\cite{Borghini:2001vi, Miller:2003kd} can be also straightforwardly computed for
a Gaussian distribution in $r$. For example 
\begin{eqnarray}
\epsilon _{2}^{2}\{2\} &=&\frac{2}{1+N},  \label{e22g} \\
\epsilon _{2}^{4}\{4\} &=&\frac{16}{(1+N)^{2}(3+N)}, \\
\epsilon _{2}^{6}\{6\} &=&\frac{192}{(1+N)^{3}(3+N)(5+N)}, \\
\epsilon _{2}^{8}\{8\} &=&\frac{6144(17+5N)}{11(1+N)^{4}(3+N)^{2}(5+N)(7+N)}.
\label{e28g}
\end{eqnarray}%
Equations (\ref{e22g}-\ref{e28g}) agree  with those obtained in
Ref. \cite{Yan:2013laa}. It is easy to verify that 
\begin{equation}
\epsilon _{2}\{2\}>\epsilon _{2}\{4\}\approx \epsilon _{2}\{6\}\approx
\epsilon _{2}\{8\}  \label{e2-2468}
\end{equation}%
in agreement with the recent numerical calculations of Ref.~\cite%
{Bzdak:2013rya}. In Figure \ref{fig:1} we compare our analytical results
with the full Monte Carlo calculations.
We checked that the small deviation from the numerical results at small $N$
comes solely from the recentering\footnote{%
Here, recentering is an event-by-event shift to the center of mass.}
correction that is neglected in our analytical calculations.\footnote{As pointed out in Ref. \cite{Yan:2013laa}, recentering can be effectively included by changing $N \rightarrow N-1$. In this case both MC and our analytical calculations agree very well for all $N$.} 

We also performed calculations for 
different functions, e.g., $P_r(r)=e^{-r^{k}/\sigma^k}$, $k>2$, as discussed in the next Section.

\section{General relations}
\label{sec:general-relations}

It is not always possible to calculate analytically $\langle I_{0}(2r^{2}%
\sqrt{z})\rangle $ and $\langle e^{-r^{2}z}\rangle $, thus in this Section
we derive general relations between $\left\langle \epsilon
_{2}^{2n}\right\rangle $ and $\left\langle r^{m}\right\rangle $ for an arbitrary function $P_r(r)$. Taking
derivatives of Eq. (\ref{e2n2deno}) and Eq. (\ref{e2n2noms}) (second line)
we obtain [$N_n \equiv \frac{N!}{(N-n)!} $ and $N \equiv N_1$] 
\begin{equation}
\left\langle \epsilon _{2}^{2}\right\rangle =\frac{{N\langle r^{4}\rangle }}{%
{N_2\langle r^{2}\rangle ^{2}+N\langle r^{4}\rangle }}\;,
\end{equation}
\begin{eqnarray}
\left\langle \epsilon _{2}^{4}\right\rangle _{\mathrm{num}}
&=&2N_2  \langle r^{4}\rangle ^{2}+N\langle r^{8}\rangle\;,  \\
\left\langle \epsilon _{2}^{4}\right\rangle _{\mathrm{deno}}
&=& N_4 \langle r^{2}\rangle ^{4}+ \notag 6 N_3 \langle r^{2}\rangle ^{2}\langle r^{4}\rangle + 3 N_2  \langle r^{4}\rangle ^{2}+ 4 N_2 \langle r^{2}\rangle \langle r^{6}\rangle + N\langle r^{8}\rangle \;,
\end{eqnarray}%
\begin{eqnarray}
\left\langle \epsilon _{2}^{6}\right\rangle _{\mathrm{num}}
&=&6  N_3\langle r^{4}\rangle ^{3}+9 N_2 \langle r^{4}\rangle
\langle r^{8}\rangle +N\langle r^{12}\rangle \;,  \\
\left\langle \epsilon _{2}^{6}\right\rangle _{\mathrm{deno}}
&=& N_6  \langle r^{2}\rangle ^{6}+  15 N_5 \langle r^{2}\rangle ^{4}\langle r^{4}\rangle +
45 N_4\langle r^{2}\rangle ^{2}\langle r^{4}\rangle ^{2}+ 
15N_3\langle r^{4}\rangle ^{3} + \notag \\ 
&&  20N_4 \langle r^{2}\rangle ^{3}\langle r^{6}\rangle + 
60N_3\langle r^{2}\rangle \langle r^{4}\rangle \langle
r^{6}\rangle +  10N_2\langle r^{6}\rangle ^{2}+  15N_3\langle r^{2}\rangle ^{2}\langle r^{8}\rangle +  \notag \\
&&15 N_2\langle r^{4}\rangle \langle r^{8}\rangle + 6 N_2\langle r^{2}\rangle \langle r^{10}\rangle + N\langle r^{12}\rangle 
\end{eqnarray}%
and finally for $\left\langle \epsilon _{2}^{8}\right\rangle $%
\begin{eqnarray}
\left\langle \epsilon _{2}^{8}\right\rangle _{\mathrm{num}}
&=&24N_4\langle r^{4}\rangle ^{4}+72N_3\langle
r^{4}\rangle ^{2}\langle r^{8}\rangle +
18N_2\langle r^{8}\rangle ^{2}+16N_2\langle r^{4}\rangle \langle
r^{12}\rangle +N\langle r^{16}\rangle \;, \\
\left\langle \epsilon _{2}^{8}\right\rangle _{\mathrm{deno}}
&=&N_8\langle r^{2}\rangle ^{8}+ 
28N_7\langle r^{2}\rangle ^{6}\langle
r^{4}\rangle +  210N_6\langle r^{2}\rangle ^{4}\langle
r^{4}\rangle ^{2}+  420N_5\langle r^{2}\rangle ^{2}\langle r^{4}\rangle
^{3}+  \notag \\
&&105N_4\langle r^{4}\rangle ^{4}+  56N_6\langle r^{2}\rangle ^{5}\langle
r^{6}\rangle +  560N_5\langle r^{2}\rangle ^{3}\langle r^{4}\rangle
\langle r^{6}\rangle +  840N_4\langle r^{2}\rangle \langle r^{4}\rangle
^{2}\langle r^{6}\rangle +  \notag \\
&&280N_4\langle r^{2}\rangle ^{2}\langle r^{6}\rangle ^{2}+ 
280N_3\langle r^{4}\rangle \langle r^{6}\rangle ^{2}+  70N_5\langle r^{2}\rangle ^{4}\langle r^{8}\rangle +420N_4\langle r^{2}\rangle ^{2}\langle r^{4}\rangle
\langle r^{8}\rangle +  \notag \\
&&210N_3\langle r^{4}\rangle ^{2}\langle r^{8}\rangle +  280N_3\langle r^{2}\rangle \langle r^{6}\rangle \langle
r^{8}\rangle +  35N_2\langle r^{8}\rangle ^{2}+  56N_4\langle r^{2}\rangle ^{3}\langle r^{10}\rangle + 
\notag \\
&&168N_3\langle r^{2}\rangle \langle r^{4}\rangle \langle
r^{10}\rangle +  56N_2\langle r^{6}\rangle \langle r^{10}\rangle +  28N_3\langle r^{2}\rangle ^{2}\langle r^{12}\rangle +  28N_2\langle r^{4}\rangle \langle r^{12}\rangle +  \notag \\
&&8N_2\langle r^{2}\rangle \langle r^{14}\rangle +  N\langle r^{16}\rangle \;.
\end{eqnarray}

Using above equations we performed calculations for various functions $P_{r}(r)$, and found 
that Eq. (\ref{e2-2468}) is always satisfied with good accuracy.
\footnote{In particular we considered $P_{r}(r)\sim e^{-r^{k}/\sigma ^{k}}$ with $\left\langle
r^{m}\right\rangle =\sigma ^{m}\Gamma (\frac{2+m}{k})/\Gamma (\frac{2}{k})$
for various values of $k$, and $P_{r}(r)\sim \Theta (r_{0}-r)$ with $%
\left\langle r^{m}\right\rangle =2r_{0}^{m}/(2+m)$.}

\section{Concluding Remarks}
\label{sec:concluding-remarks}

In conclusion, we find an exact formula for $\left\langle \epsilon
_{2}^{2n}\right\rangle $ for $N$ independent points sampled according to a
general distribution $P(r,\phi )$. We restrict ourselves to a systems
dominated by fluctuations, so that on average this system is
azimuthally symmetric.  This should apply to p+A collisions at the LHC.
We also derived
explicit relations for the cumulants, $\epsilon _{2}\{2n\}$, for a Gaussian
distribution. Finally, we analytically verified the recently observed
numerical relation $\epsilon _{2}\{2\}>\epsilon _{2}\{4\}\simeq \epsilon
_{2}\{{6\}}\simeq \epsilon _{2}\{8\}$ between cumulants in p+A collisions.

\section*{Acknowledgments}
\hspace*{\parindent} 
We thank R. Pisarski for valuable comments. 
A.B. was supported through the RIKEN-BNL Research
Center, by the Ministry of Science and Higher Education (MNiSW), by founding from the Foundation for Polish Science, and by the National Science Centre, Grant No. DEC-2014/15/B/ST2/00175, and in part by DEC-2013/09/B/ST2/00497. V.S. acknowledges support and hospitality of RIKEN-BNL Research Center, where 
this work was initiated.

\end{document}